# Structural and magnetic study of a dilute magnetic semiconductor: Fe doped $CeO_2$ nanoparticles

Shalendra Kumar[1]*, Geun Woo Kim[1], B. H. Koo[1], S. K. Sharma[2], M. Knobel[2] and Chan Gyu Lee[1]†

[1]School of Nano and Advanced Materials Engineering, Changwon National University, # 9 Sarim-dong, Changwon- 641-773, Republic of Korea

[2]Instituto de Fisica Gleb Wataghin, Universidade Estadual de Campinas, (UNICAMP) Campinas, 13.083-970, SP, Brazil

This paper reports the effect of Fe doping on the structure and room temperature ferromagnetism of $CeO_2$ nanoparticles. X-ray diffraction and selective area electron diffraction measurement reflects that $Ce_{1-x}Fe_xO_2$ ($0 \leq x \leq 0.07$) nanoparticles exhibit single phase nature with cubic structure and none of the sample showed the presence of any secondary phase. The mean particle size calculated by using a transmission electron microscopy measurement was found to increase with increase in Fe content. DC magnetization measurements performed at room temperature indicates that all the samples exhibit ferromagnetism. The saturation magnetic moment has been found to increase with an increase in the Fe content.

Keywords: DMS; $CeO_2$; X-ray diffraction; FTIR, DC magnetization, RT-FM

Corresponding author

*E-mail: shailuphy@gmail.com (Shalendra Kumar);

†E-mail: chglee@changwon.ac.kr (C. G. Lee);

Ph: +82-55-213-3703; Fax: +82-55-261-7017

1. INTRODUCTION

Dilute magnetic semiconductors (DMSs) have drawn considerable attention in the past few years in view of their projected potential for the development of the magneto-optoelectronics devices.[1-3] It is generally accepted that ferromagnetism in semiconductors requires a few percent of the transition metal (TM) ions dopants that have partially filled shell of d or f electrons to mediate ferromagnetism.[4] However, certain DMS systems based on III-V and II-VI semiconductors appear to have been recognized as to have intrinsic (carrier-induced) ferromagnets[5], but the situation with the oxide based system is still controversial, many promising compounds such as: transition metal (TM) doped ZnO, $TiO_2$, $CeO_2$ and $HfO_2$ etc, have been observed to show room temperature ferromagnetism (RT-FM). However, for most of the experimental results, doubt arose about the real origination of the FM.[6-11] Some of the report claims that the observed FM is induced by the segregation of metallic clusters or due to the secondary phase, while in some systems it is due to the different valence state of TM ions. Although, the high $T_C$ FM observed in TM doped insulating oxides has been explained using different models even in the same system by the different groups. In the early stage, the origin of FM in oxide based DMSs were explained by the carrier mediated exchange.[5] Recently, Croy et al[10] have been proposed donor impurity band exchange in DMSs. According to this model ferromagnetic exchange in DMSs are mediated by the shallow donor electrons that form bound magnetic polarons, which overlap to create a spin-split impurity band. More recently, the origin of high temperature FM in insulating oxides has been explained by the oxygen vacancy mediated FM or so-called F-centre exchange coupling (FCE) mechanism.[12] In FCE mechanism an electron is trapped in oxygen vacancy which acts as a coupling centre, via which doped magnetic ions align in ferromagnetic order.

In the last few years, $CeO_2$ containing materials have received a lot of attention due to its diversity of applications such as: catalysis[13], in solid oxide fuel cell[14], in ceramic materials[15] and in spintronics.[16-20] $CeO_2$ nanoparticles have been prepared using several methods such as hydrothermal synthesis[21], hexamethylenetetramine based homogeneous precipitation[22], co-precipitation[23] and force hydrolysis[24] etc. However, there are copious methods to synthesize $CeO_2$ nanoparticles, but a very small work has been carried out to investigate the effect of the TM doping on the magnetic properties of $CeO_2$. The discovery of FM in $CeO_2$ tiles the way for the realization spintronics devices. $CeO_2$ is an insulating material with high dielectric constant (~26). Since $CeO_2$ has FCC structure like fluoride crystal structure and closely lattice matches with silicon, therefore, the RT-FM in $CeO_2$ will make possible the integration spintronics devices with the advanced silicon based microelectronics. In the present work, we have prepared Fe doped $CeO_2$ nanoparticles using co-precipitation technique. The effect of Fe doping has been studied using x-ray diffraction (XRD), transmission electron microscopy (TEM), Fourier transform infrared spectroscopy (FTIR) and DC magnetization measurements. The results of XRD, FTIR and TEM measurements clearly indicate that Fe ions are substituting Ce ions in the $CeO_2$ matrix. The DC magnetization measurement also confirms that all the samples display RT-FM.

2. EXPERIMENTAL DETAILS

Nanoparticles of Fe doped $CeO_2$ have been synthesized using the co-precipitation method. Highly pure analytical grade metal nitrate [$Ce(NO_3)_2 \cdot 6H_2O$, $Fe(NO_3)_3 \cdot 9H_2O$] were dissolved in de-ionized water to get 0.6 M solution. This solution was kept for 1 h for stirring at 25 °C. In this solution, 5 M $NH_4OH$ solution was added drop wise until the pH value reached 9. The solution was stirred for 3 h at room temperature and filtered. The mixture was dried at 100 °C for 15 h.

The dried mixture was ground and annealed at 500 $^{o}$C for 12 h in air. The detailed characterization of the sample was carried out using TGA, XRD, TEM, FTIR and DC magnetization measurements. XRD measurement was carried out using a Phillips X'pert (MPD 3040) x-ray diffractometer with a Cu K$_\alpha$ source ($\lambda$=1.5406Å) operated at voltage 36 kV and current 30 mA. TEM measurements were performed using FE-TEM (JEM 2100F). FTIR measurements were performed using Nicolet Impact 410DSP spectrometer. Thermal gravimetric (TG) measurements were performed using 2960 SDT instrument. DC Magnetization measurements were performed at room temperature using a commercial Quantum Design Physical properties measurement system (PPMS).

3. RESULTS AND DISCUSSIONS

Thermal gravimetric analysis (TGA) measurements of Fe doped $CeO_2$ samples were performed to investigate the most suitable calcination temperature. The precipitate obtained after washed by the distilled water and dried at 100 $^{0}$C needs the thermal treatment to form the single phase metal oxide powder. Fig. 1 presents the TGA curve of 1 and 7% Fe doped $CeO_2$ nanoparticles, heated at rate of 10 $^{o}$C/min up to 1000 $^{o}$C in air. It can be seen from Fig. 1 that the weight loss decreases with increase in the temperature upto ~ 500 $^{o}$C. After 500 $^{o}$C, there is no appreciable change in the weight loss, therefore, we can say that the suitable calcination temperature is at least over 500 °C. The colors of the as synthesized samples are somewhat yellowish.

Fig. 2 highlights the XRD pattern of $Ce_{1-x}Fe_xO_2$ (x = 0.01, 0.05 and 0.07) nanoparticles. It is clear from the XRD pattern that all the samples show single phase nature with fluorite cubic structure. None of the sample showed any extra peak corresponding to any metallic cluster or oxide phase of Fe, therefore, exclude the presence of secondary phase. Moreover, the broadening of the XRD peaks clearly indicates that the prepared samples have nanocrystalline behavior. It can be seen from the XRD pattern that the peaks position have shift towards the higher 2θ value with increase in Fe content. Inset (a) in Fig. 2 represents the d values calculated using (111) peak for different concentration of Fe ions doped in $CeO_2$ matrix. The d (111) values are found to decreases with the Fe content. Thus, the decreasing trend of d (111) values reflects that lattice parameters decreases with Fe doping which is in good agreement with the earlier reported results.[16, 25, 26] This behavior is expected due to the difference in the ionic radii of cerium and iron. Therefore, the XRD results clearly indicate that Fe ions are replacing Ce ions in $CeO_2$ matrix and ruled out the formation of any extra phase.

Fig. 3(a) and (b) represents the TEM micrograph, selected area electron diffraction (SAED) pattern and particle size distribution histogram of $Ce_{1-x}Fe_xO_2$ (x = 0.01 and 0.07) nanoparticles. TEM micrographs indicate that the nanoparticles have spherical shape. The particle size distribution obtained from TEM micrographs using Image J 1.3.2 J software reflects that particle size increases with Fe doping. It is observed that 1 and 7% Fe doped $CeO_2$ nanoparticles have particle size distribution from 3.5 - 8 and 4 - 12 nm respectively. The mean particle size calculated for 1 and 7% Fe doped $CeO_2$ nanoparticles from the fitting of the particle distribution histogram using Lorentzian function are found 5.7 and 7.5 nm respectively. SAED measurement were performed to further study the single phase nature of Fe doped $CeO_2$ nanoparticles, since, it is difficult to detect any impurity phase below 5% using XRD measurements. Inset in Fig. 3(a) and (b) shows the SAED pattern of Fe doped $CeO_2$ nanoparticles. Rings in SAED patterns indicate clearly the randomly oriented single crystals, hence ruled out the presence of any impurity phase and indicate that each nanoparticle is indeed in single phase.

FTIR measurement of Fe doped $CeO_2$ nanoparticles carried out in the wave number range 2000 to 400 $cm^{-1}$ is shown in Fig. 4. The FTIR measurements been performed to investigate the effect of Fe doping on the Ce-O-Ce bonding. FTIR measurements were done using KBr method at RT. The absorption band observed around 1625 cm$^{-1}$ is attributed to the stretching mode hydroxyl.[27] The peak at 846 cm$^{-1}$ is due to the vibrations associated with the incordination of the absorbed $NO_3^{-1}$ ions[28]. The spectra of each samples show the absorption bands at about 1050 and 1320 cm$^{-1}$ which is the characteristic vibrations of $CeO_2$.[29]

Fig. 5 displays magnetization (M) versus magnetic field curve of Fe doped $CeO_2$ samples at room temperature. The value of coercivity ($H_C$), remanent magnetization ($M_r$) and saturation magnetization ($M_s$) calculated from M-H curve is shown in table. The value of $H_C$ has been found to decrease from 66.7 to 28.6 Oe where as the values $M_r$ and $M_S$ have been found to increase from $4.2\times10^{-3}$ to $5.2\times10^{-3}$ emu/g and $5.5\times10^{-3}$ to $14\times10^{-3}$ emu/g respectively with increase in the content of Fe in $CeO_2$ matrix (see table). It can be clearly seen from the inset in Fig. 5 that all samples exhibit RT-FM. Earlier in case of Co doped $CeO_2$ nanoperticles[25], we have observed that saturation magnetization decreases with increase in Co doping beyond 3%. Thurber el al[26] also reported that both the saturation magnetization and remanent magnetization increases with Ni doped $CeO_2$ nanoparticles up to 4% and then decreases with further increase in the Ni doping. The magnetic ordering in the insulating materials is generally originated by the super-exchange interaction, which results antiferromagnetic ordering. However, recently another conceivable mechanism has been proposed to explain the FM in transparent oxide is FCE mechanism.[12] According to the FCE mechanism, the magnetic ion and oxygen vacancy are two necessary requirements which results the FM ordering. The FM observed in the present system can be explained in the light of the FCE mechanism. In FCE mechanism an F centre consists of an electron trapped in the oxygen vacancy. The oxygen vacancy constitutes groups with two Fe ions (i.e. Fe-$O_v$-Fe, where $O_v$ represent oxygen vacancy). An electron trapped in oxygen vacancy occupies an orbital which overlaps the d shell of neighboring Fe ions. According to the Hund's rule and Pauli principle the trapped electrons spin should have direction parallel to two neighboring Fe ions, which results FM ordering. The complex magnetic behavior observed in TM doped $CeO_2$ nanoparticles can be explained as follows: As the TM ions are doped in $CeO_2$ matrix, they are mediated by FCE interactions due to the small separation between them, which leads increases in the magnetization with TM contents. However, with further increase in TM ions may results super-exchange interactions, as a result the saturation magnetization decreases due to the antiferromagnetic ordering. Though, in case of Fe doped $CeO_2$ the magnetization increases with Fe doping which may be due to the mix valance state of Fe ions.

4. CONCLUSIONS

We have successfully prepared the nanoparticles of Fe doped $CeO_2$ using co-precipitation technique. TGA measurements show that the most suitable calcination temperature is ~ 500 $^oC$. XRD and SAED results indicate that all the samples have single phase nature with cubic structure. The shifting of the XRD peaks and thus decrease in the d value with increase in the Fe content clearly indicates that Fe ions are replacing Ce ions in $CeO_2$ matrix and excludes the possibility of any secondary phase formation. DC magnetization measurements reveal that all the samples exhibit RT-FM and the saturation magnetization increases with Fe content in $CeO_2$ matrix.

Acknowledgements

This work was supported by the Korean Research Foundation Grant funded by the Korean Government (MOEHRD) (KRF-2006-005-J02703).

References

1. H. Ohno Science 281, 951 (1998)

2. G. A. Prinz, Science 282, 1660 (1998)

3. S. D. Sharma, Am. Sci. 89, 516 (2001)

4. T. Dietl, H. Ohno, F. Matsukura, J. Cibert, and D. Ferrand, Science 287, 1019 (2000)

5. H. Ohno, A. Sen, F. Matsukura, A. Oiwa, A. Endo, S. Katsumoto. Y. lye, Appl. Phys. Lett. 69, 363 (1996)

6. H. Ohno, J. Magn. Magn. Mater. 200, 110 (1999)

7. M. Bouloudenine, N. Viart, S. Colis, J. Kortus, and A. Dinia, Appl. Phys. Lett. 87, 052501 (2005)

8. John E. Jaffe, Timothy C. Droubay, Scott A. Chambers, J. Appl. Phys. 97, 073908 (2005)

9. H. Weng, X. Yang, J. Dong, H. Mizuseki, M. Kawasaki, Y. Kawazoe, Phys. Rev. B 69, 125219 (2004).

10. J. M. D. Coey, M. Venkatesan, C. B. Fitzgerald, Nat. Mater. 4, 173 (2005)

11. M. Venkateshan, C. B. Fitzgerald, J. M. D. Coey, Nature, 430, 630 (2004)

12. J. M. D. Coey, A. P. Douvalis, C. B. Fitzgerald, M. Venkatesan, Appl. Phys. Lett. 84 1332 (2004)

13. S. Rossignol, Y. Madier, D. Duprez, Catal. Today 50, 173 (1999)

14. M. Shiono, K. Kobayashi, T.L. Nguyen, K. Hosoda, T. Kato, K. Ota, M. Dokiya, Solid State Ionic 170, 1 (2004)

15. A. Trovarelli, C.D. Leitenburg, G. Dolcetti, Chem. Tech. 27, 32 (1997)

16. Qi.Y. Wen, H. W. Zhang, Y.Q. Song, Q.H. Yang, H. Zhu, J.Q. Xiao, J. Phys : Condens Matter. 19, 246205 (2007)

17. Y.Q.Song, H.W. Zhang, Q.Y. Wen, L. Peng, J. Q. Xiao, J. Phys : Condens Matter. 20, 255210 (2008)

18. Y.Q.Song, H.W. Zhang, Q.Y. Wen, H. Zhu, J. Q. Xiao, J. Appl. Phys. 102, 043912 (2007)

19. V. Fernandes, J.J. Klein, N. Mattoso, D.H. Mosca, E. Silveira, E. Ribeiro, W.H. Schreiner, J. Varalda, A.J.A. de Oliveira, Phy. Rev. B 75, 121304(R) (2007)

20. B. Vodungbo, Y. Zheng, F. Vidal, D. Demaille, V.H. Etgens, D.H. Mosca, Appl. Phys. Lett. 90, 062510 (2007)

21. M. Hirano, E. Kato, J. Am. Ceram. Soc. 79, 777 (1996)

22. P.L Chen, I. W, Chen, J. Am. Ceram. Soc. 79, 3129 (1996)

23. X. Chu, W. Chung, L. D. Schmidt, J. Am. Ceram. Soc. 76, 2115 (1996)

24. X. D. Zhou, W. Huebner, H.U. Anderson, Appl. Phys. Lett. 80, 3814 (2002)

25. Shalendra Kumar, Y.J. Kim, B.H. Koo, Chan Gyu Lee, IEEE Trans. Magn. 45, 2439 (2009)

26. A. Thurber, K. M. Reddy, V. Shutthanandan, M. H. Engelhard, C. Wang, J. Hays, and A. Punnoose, Phys. Rev. B 76, 165206 (2007)

27. J. Rubio, J. L. Oteo, M. Villegas, P. Duran, J. Mater. Sci. 32, 643 (1997)

28. Y. W. Zhang, S. Rui, C. S. Liao, C. H. Yan, J. Phys. Chem. B 107, 10159 (2003)

29. B. Yan, H. Zhu, J. Nanopart. Res. 10, 1279 (2008)

Figure captions

Fig. 1. TGA curve for $Ce_{1-x}Fe_xO_2$ (x = 0.01 and 0.07).

Fig. 2. (Color online) XRD pattern for $Ce_{1-x}Fe_xO_2$ (x = 0.0, 0.01, 0.05 and 0.07). Inset (a) shows the d value as a function of Fe content.

Fig. 3. TEM images for the $Ce_{1-x}Fe_xO_2$ and the corresponding particle size histograms for (a) x = 0.01 and (b) x = 0.07. Inset shows the corresponding SAED pattern and particle size histograms.

Fig. 4. (Color online) FTIR spectra of $Ce_{1-x}Fe_xO_2$ (x = 0.01, 0.05 and 0.07) nanoparticle samples taken at room temperature.

Fig. 5. (color online) M versus H curves taken at room temperature for nanoparticles of $Ce_{1-x}Fe_xO_2$ (x = 0.01, 0.05 and 0.07). Inset showing the expanded view of low field region.

Table caption

Table 1 Calculated value of coercive field ($H_C$), remanent magnetization ($M_r$) and saturation magnetization ($M_S$) for $Ce_{1-x}Fe_xO_2$ (x = 0.01, 0.05 and 0.07)

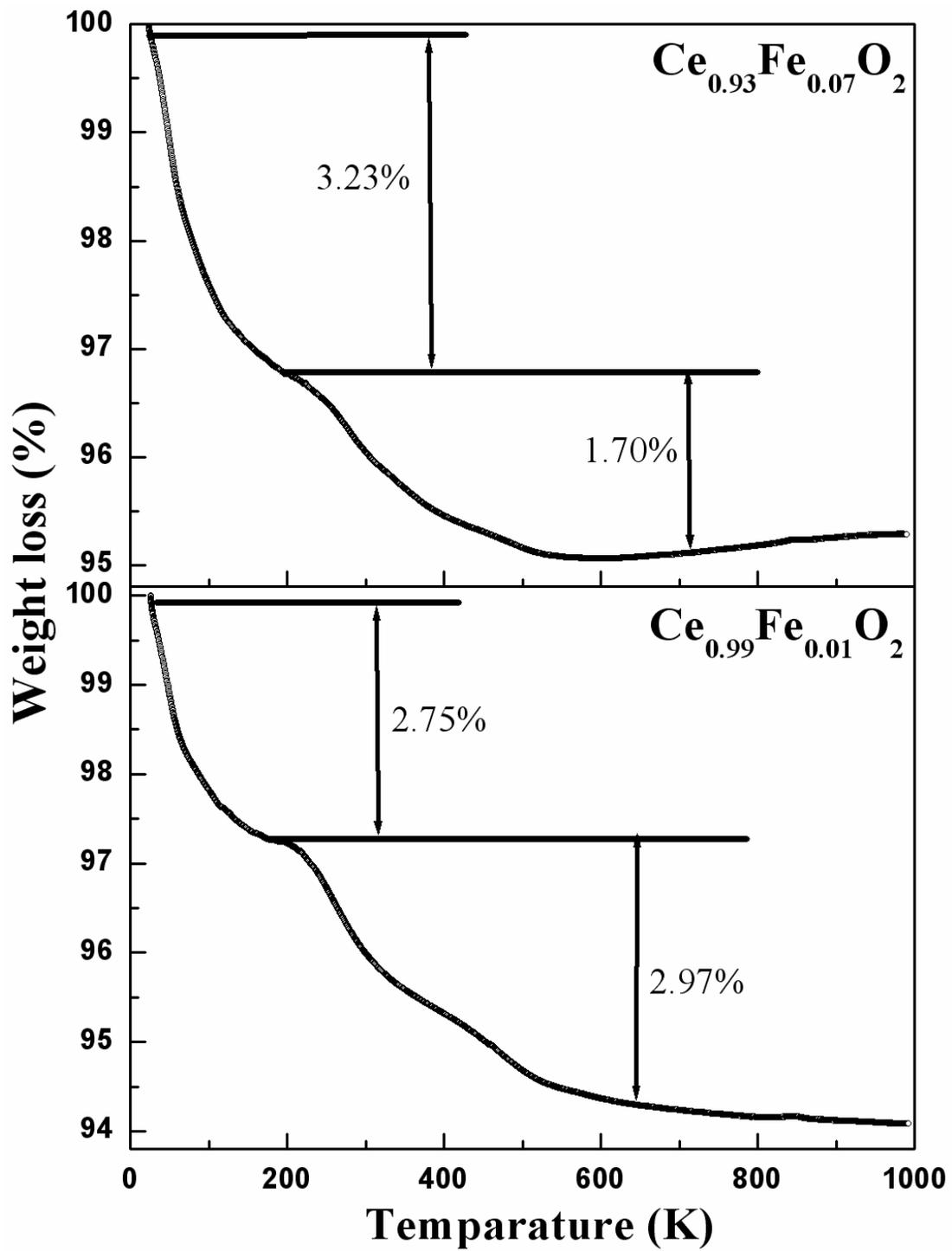

Fig. 1

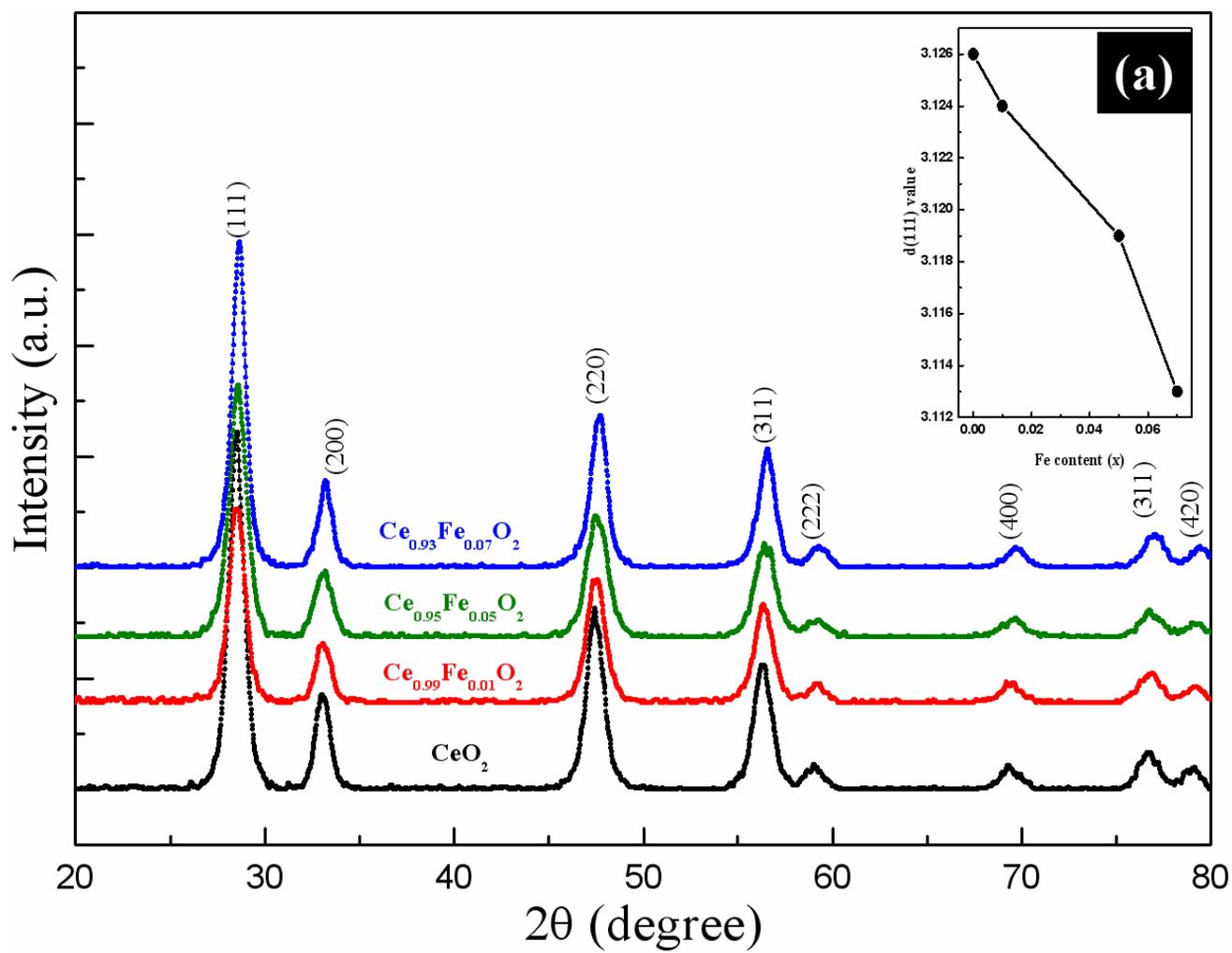

Fig. 2

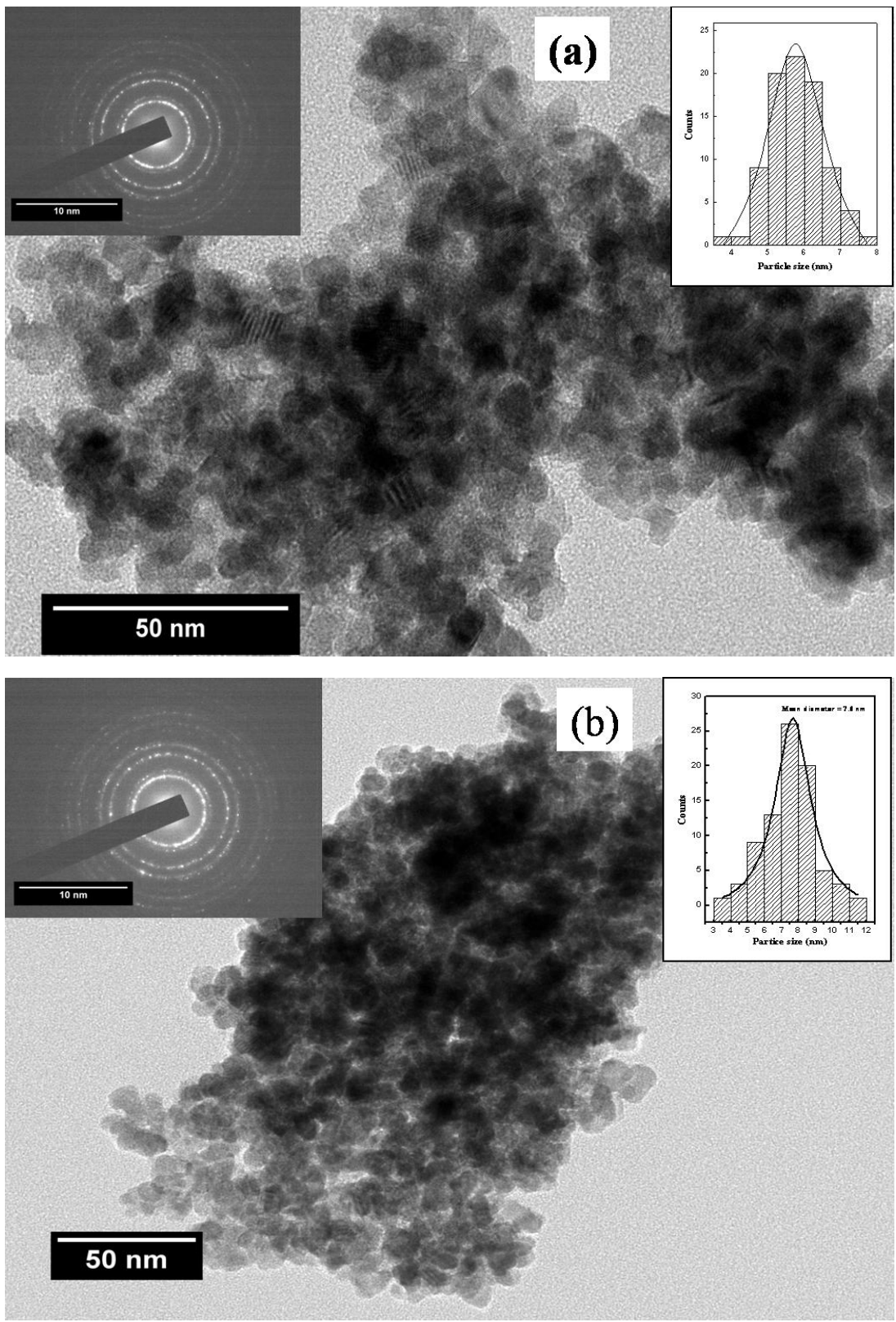

Fig. 3 (a) and (b)

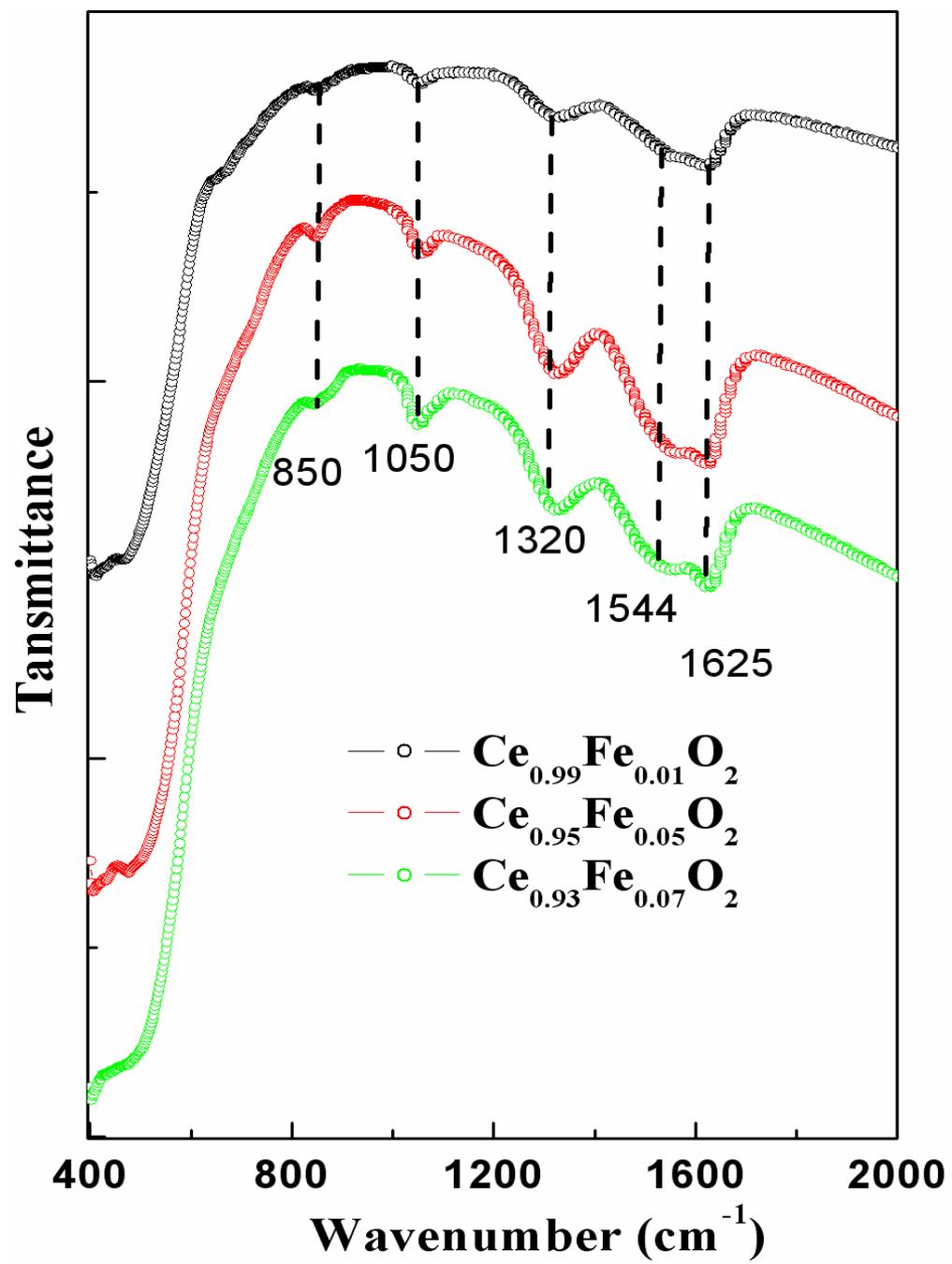

Fig. 4

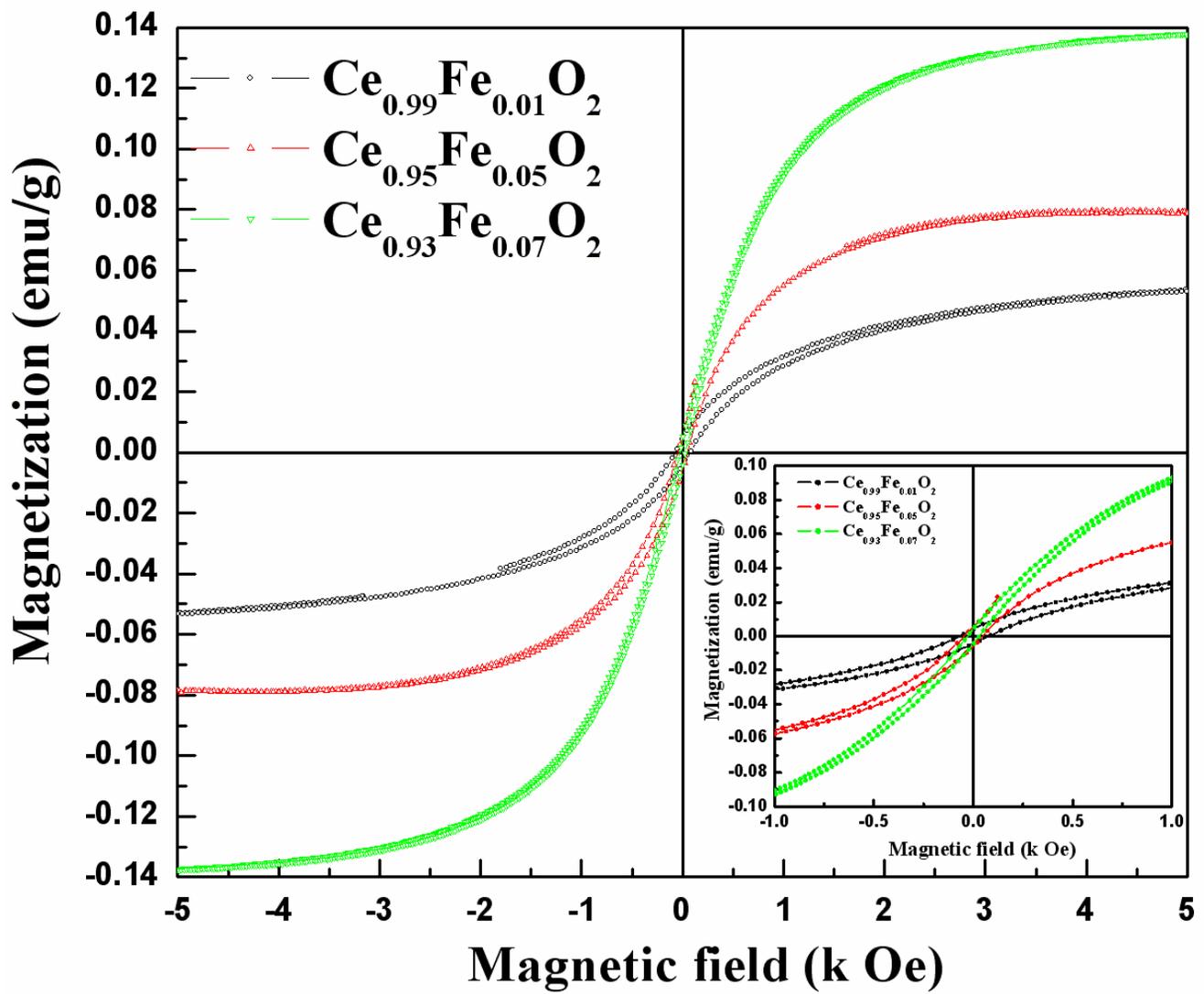

Fig. 5

Table 1

| Sample | Coercive field ($H_C$) (Oe) | Remanent magnetization ($M_r$) (emu/g) $\times 10^{-3}$ | Saturation magnetization ($M_S$) (emu/g) $\times 10^{-3}$ |
|---|---|---|---|
| $Ce_{0.99}Fe_{0.01}O_2$ | 66.7 | 4.2 | 5.5 |
| $Ce_{0.95}Fe_{0.05}O_2$ | 46.5 | 5.0 | 7.9 |
| $Ce_{0.93}Fe_{0.07}O_2$ | 28.6 | 5.2 | 14.0 |